\documentclass[sigconf,screen]{acmart}
\usepackage{booktabs}
\usepackage{array}
\usepackage{xcolor}
\usepackage{graphicx}
\usepackage{enumitem}
\newcommand{\ATIBA}{ATIBA}

\setcopyright{none}
\setcctype{by-nc-nd}
\acmDOI{10.1145/3844135.3845865}
\acmYear{2026}
\copyrightyear{2026}
\acmISBN{979-8-4007-2999-7/2026/10}
\acmConference[ATIQSER '26]{Proceedings of the 1st International Workshop on Automated Techniques for Integrity and Quality in Software-Engineering Research}{October 12--16, 2026}{Munich, Germany}
\acmBooktitle{Proceedings of the 1st International Workshop on Automated Techniques for Integrity and Quality in Software-Engineering Research (ATIQSER '26), October 12--16, 2026, Munich, Germany}
\acmSubmissionID{asews26atiqsermain-p2-p}
\received{2026-08-16}
\received[accepted]{2026-09-01}

\begin{document}

\title{ATIBA: Grounded Integrity and Quality Checking for Research Papers}

\author{Veli Karakaya}
\authornote{These authors contributed equally to this work.}
\orcid{0009-0009-5389-6588}
\affiliation{%
  \institution{Bilkent University}
  \department{Computer Engineering}
  \city{Ankara}
  \country{Turkiye}
}
\email{veli.karakaya@ug.bilkent.edu.tr}

\author{Semih Çağlar}
\authornotemark[1]
\orcid{0009-0003-1311-5747}
\affiliation{%
  \institution{Bilkent University}
  \department{Computer Engineering}
  \city{Ankara}
  \country{Turkiye}
}
\email{semih.caglar@ug.bilkent.edu.tr}

\author{Yusuf Yiğit Korkmaz}
\authornotemark[1]
\orcid{0009-0006-2320-7399}
\affiliation{%
  \institution{Bilkent University}
  \department{Computer Engineering}
  \city{Ankara}
  \country{Turkiye}
}
\email{yigit.korkmaz@ug.bilkent.edu.tr}

\author{Eray Tüzün}
\correspondingauthor
\orcid{0000-0002-5550-7816}
\affiliation{%
  \institution{Bilkent University}
  \department{Computer Engineering}
  \city{Ankara}
  \country{Turkiye}
}
\email{eraytuzun@cs.bilkent.edu.tr}

\renewcommand{\shortauthors}{Karakaya et al.}

\begin{abstract}
Checking a manuscript's reference integrity, its compliance with a target
venue's specific submission rules, and its adherence to community reporting
standards is manual, repetitive, and different for every venue so in
practice it is done inconsistently or skipped. We present \ATIBA{}, a tool
that runs five grounded integrity and quality checks on a manuscript: a
reference-integrity check that verifies each citation against bibliographic
sources and flags retracted or unfindable references; a venue/track
compliance check that derives submission criteria directly from a venue's own
call-for-papers page and evaluates the manuscript against them, each verdict
anchored to a verbatim quote from that page; an empirical-standards
compliance check against the ACM SIGSOFT Empirical Standards, with a
hallucination defence that discards any evidence quote it cannot locate
verbatim in the manuscript; a multi-mode AI review (venue-specific, formal, and page-anchored annotation) powered by GPT-5.4 through Azure OpenAI; and a citation-suggestion feature that proposes candidate references for a manuscript and verifies each against bibliographic sources before it is shown to the user. All five checks are designed around the
same principle: an LLM is only trusted to judge, never to invent the evidence
it judges against. We evaluated \ATIBA{} through a moderated user study with
13 non-author participants. Agreement across the six survey items ranged from
69\% to 92\%, with a mean of 85\%,
providing initial evidence of positive perceived usefulness across the
evaluated workflows. These findings establish perceived usefulness; objective accuracy remains to be measured.
\end{abstract}

\keywords{Research integrity, Research quality, Automated peer review, Reference hallucination, Large language models, Venue compliance}

\ccsdesc[500]{Applied computing~Publishing}
\ccsdesc[300]{Software and its engineering}
\ccsdesc[100]{Social and professional topics~Codes of ethics}

\maketitle

\section{Introduction}

Submission counts at top-tier computing venues have grown sharply in recent years, placing increasing strain on traditional peer review~\cite{biswas2026aiassistedpeerreviewscale, jayaram2026automatingscientificreviewgoogles}, a trend also reflected in ASE 2025 submission data~\cite{xoveexu2026csconfstats}. Reviewing a submission for integrity and quality problems means checking several different things at once: are the references real and verifiable~\cite{huang2023hallucination,walters2023fabrication}, have any of them been retracted, does the manuscript follow the target venue's specific
formatting and anonymization rules, does it report its methodology the way
the community expects, and is the contribution itself sound. Authors are expected to check relevant submission requirements before submitting, while venues perform editorial screening before manuscripts are sent for external review \cite{goldberg2024usefulnessllmsauthorchecklist, Lin_2023}. These checks span formatting, scope, integrity, and other quality criteria, and the applicable requirements vary across publication venues \cite{Lin_2023}. In practice, carrying them out consistently imposes substantial effort, particularly because each venue defines its own aims, review standards, and publication requirements.

Existing tooling addresses pieces of this problem but not the combination.
Conference management systems such as HotCRP~\cite{hotcrp} and
EasyChair~\cite{easychair} route submissions and assignments but perform no
content-level integrity checking at all. General-purpose large language model (LLM) assisted review
tools can produce plausible-sounding feedback on a manuscript~\cite{liang2023llmfeedback},
but an LLM asked to judge a paper's references or its compliance with a venue's
rules is exactly as capable of inventing a plausible-sounding but false
verdict as it is of catching a real problem~\cite{huang2023hallucination}.

\ATIBA{}\footnote{\ATIBA{} is deployed and publicly available at
\url{https://atiba.onrender.com}.} is built around a different default:
every one of its five checks is grounded in something outside the LLM's own
output, such as a bibliographic lookup, a verbatim quote from the venue's own
page, or a verbatim quote from the manuscript itself. This grounding allows a
check to fail loudly, by returning no verdict or an explicitly low-confidence
one, instead of fabricating an answer. The same
manuscript, run once, gets a reference-integrity report, a venue-specific
compliance report, an empirical-standards report, one of three styles of
AI-generated review, and a set of verified candidate citations, all cached
and re-runnable on demand.

This paper makes the following contributions:
\begin{enumerate}
  \item \textbf{Grounded Integrity Pipeline:} A multi-stage architecture that automates reference-integrity verification, venue/track compliance extraction, and empirical-standards checking against the ACM SIGSOFT Empirical Standards~\cite{ralph2020empirical}.
  \item \textbf{Verbatim Hallucination Defence:} A strict grounding mechanism that automatically discards LLM-generated feedback or compliance verdicts whenever supporting quotes cannot be located verbatim in the manuscript or venue source text.
  \item \textbf{Context-Aware Citation \& Multi-Mode Review:} A verified citation-suggestion engine and a venue-aware prompting layer providing three complementary AI review modes---venue-specific, formal, and page-anchored annotated review.
  \item \textbf{Empirical Evaluation:} A moderated user study ($N=13$) demonstrating high perceived usefulness across all five checks in real-world paper writing and internal peer-review workflows (Section~\ref{sec:evaluation}).
\end{enumerate}

\section{Background and Related Work}

\subsection{Peer-Review Tooling}
Conference management platforms such as HotCRP~\cite{hotcrp} and
EasyChair~\cite{easychair} handle submission routing and reviewer assignment,
leaving all content-level integrity and compliance checking to human authors,
reviewers, and chairs. AI-based review support has also been piloted directly
within a venue's own review pipeline, as at the AAAI-26 AI Review Pilot, where
every AAAI-26 main-track submission received a system-generated review
alongside its human reviews~\cite{biswas2026aiassistedpeerreviewscale}. \ATIBA{}'s AI review mode instead runs before submission, on the
author's side, alongside four independently grounded checks -- reference
integrity, venue compliance, empirical standards, and citation suggestion --
that neither conference management platforms nor a reviewer-facing pilot
provides.

\subsection{Empirical Standards}
The ACM SIGSOFT Empirical Standards project~\cite{ralph2020empirical} defines,
per research methodology (controlled experiment, case study, repository
mining, and others), the essential, desirable, and extraordinary attributes a
rigorous study of that type should report, replacing the piecemeal,
methodology-specific guidance that preceded
it~\cite{jedlitschka2005reporting,runeson2009guidelines}. Today, checking a
manuscript against these standards is a manual read-through.\ATIBA{}'s
empirical-standards check is the first to automate this
specific checklist while grounding every verdict in a manuscript-verbatim
evidence quote.

\subsection{Reference and Citation Verification}
Verifying that a cited work exists and is not retracted is a
bibliographic-lookup problem distinct from judging whether it is cited
appropriately. Fabricated references are not a hypothetical risk: large-scale
studies of LLM-generated literature reviews have found that a substantial
share of citations are entirely fabricated~\cite{walters2023fabrication}, and
dedicated verification tools have begun to emerge in
response~\cite{abbonato2026checkifexist}. \ATIBA{}'s reference-integrity check
delegates existence verification to a dedicated third-party tool,
\texttt{hallucinator-cli}~\cite{hallucinatorcli}, invoked as an external
subprocess. A similar strategy appears in CheckIfExist~\cite{abbonato2026checkifexist}, which cascades queries across the same bibliographic sources \ATIBA{} uses to verify its own citation suggestions (Section~\ref{sec:citationsuggestion}). \ATIBA{} additionally adds its own
LLM-based check for citation context -- whether the citation is used in the
right context -- which existence-only verification cannot answer.

\subsection{Citation Suggestion}
Context-aware citation recommendation has long aimed to identify suitable
references from the textual context surrounding a citation location. Jeong
et al.~\cite{jeong2019contextawarecitationrecommendationmodel} rank candidate
papers for a citation placeholder using BERT-based context and
citation-graph representations, and Jebari et al.~\cite{Jebari2023} survey
this broader line of local citation recommendation. More recently, Seo et
al.~\cite{Seo2026} propose a retrieval-augmented approach matching manuscript
sentences against a pre-collected reference repository, showing
substantially higher accuracy than direct LLM prompting, which suffers from
severe hallucination. \ATIBA{} shares this goal but takes a different
integrity-oriented approach: an LLM proposes candidate references, and each
is independently checked against Crossref~\cite{CrossrefAPI}, Semantic
Scholar~\cite{SemanticScholarAPI}, and OpenAlex~\cite{OpenAlexAPI} before
being shown to the user.

\subsection{LLM-Assisted Review and Hallucination Risk}
LLMs can produce manuscript feedback that meaningfully overlaps with human
reviewer comments~\cite{liang2023llmfeedback}, but LLM-based scholarly
systems remain vulnerable to unsupported or fabricated
evidence~\cite{huang2023hallucination,ju2026wispaperaischolarsearch}. WisPaper
mitigates this by requiring verbatim source-text evidence for per-criterion
judgments~\cite{ju2026wispaperaischolarsearch}, while dedicated
review-generation systems such as ReviewAgent~\cite{garg2025revieweval} and
ReviewerToo~\cite{sahu2025reviewertoo} pursue conference-aligned or
persona-based feedback without the same evidence requirement. A parallel
ecosystem of commercial pre-submission tools has emerged: PaperReview.ai
grounds its critique in retrieved arXiv papers~\cite{paperreviewai},
PapeReview runs a comparable agentic workflow~\cite{papereviewcom}, and Review-it pairs feedback with a journal-recommendation
feature~\cite{reviewitai}, suggesting where to submit instead of checking
compliance with a venue already chosen (Section~\ref{sec:venue}). \ATIBA{} follows the same
evidence-grounding principle as WisPaper but enforces it at inference time --
any supporting quote not locatable verbatim in its source is discarded
together with its verdict -- and, unlike the systems above, treats AI review
(Section~\ref{sec:review}) as one of five independently grounded checks, not the tool's sole output.
\section{Proposed System: \ATIBA{}}

\begin{figure*}[t]
    \centering
    \includegraphics[width=\linewidth]{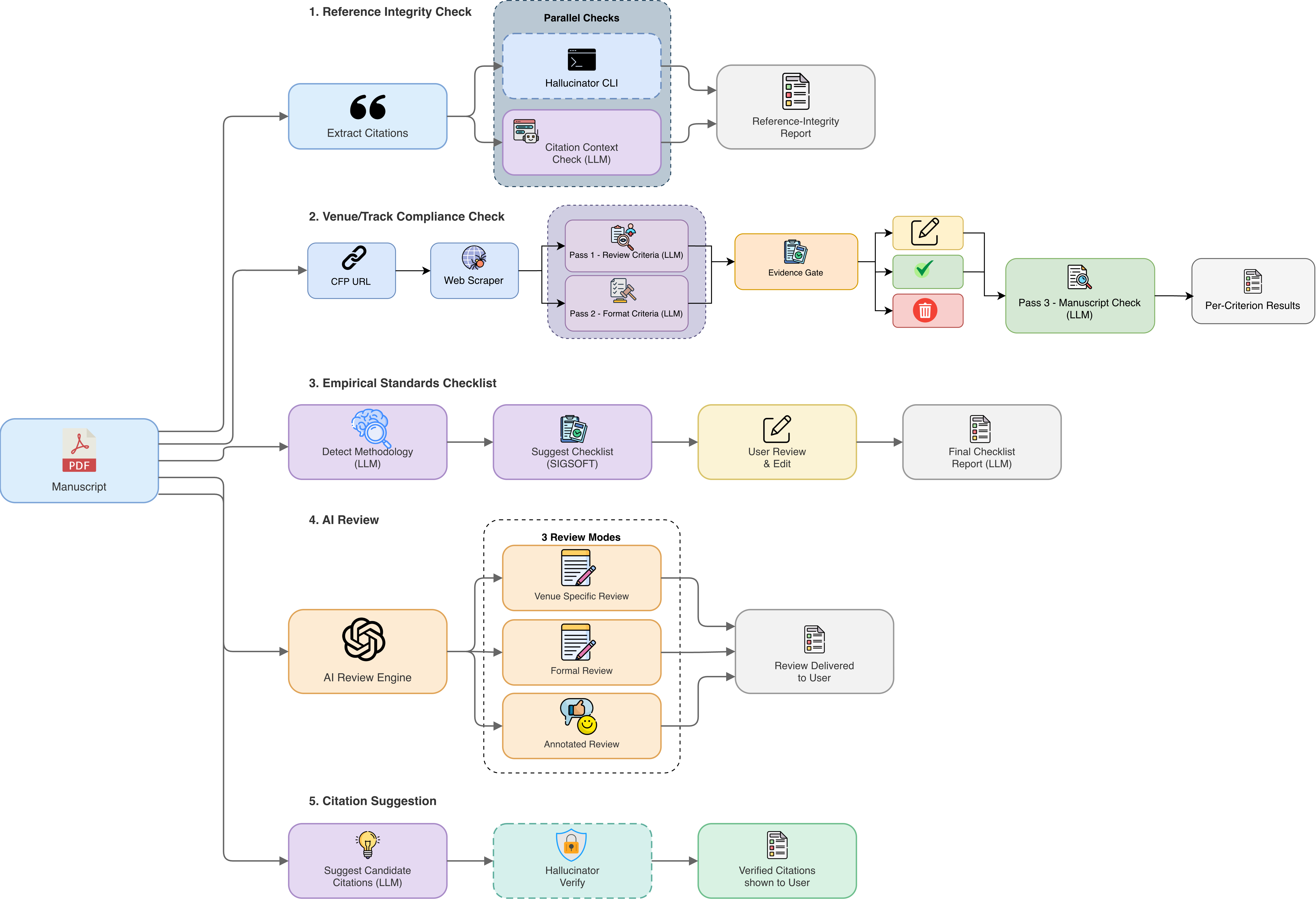}
   \caption{\ATIBA{}'s five manuscript-analysis workflows: (1) reference-integrity
verification, (2) venue/track compliance checking, (3) empirical-standards
checklist, (4) AI-assisted review in three modes, and (5) verified citation
suggestion.}
\label{fig:pipeline}
\Description{Flow diagram showing five workflows that begin with an uploaded
manuscript: (1) bibliographic and citation-context verification, (2) extraction,
user validation, and manuscript-level checking of venue criteria, (3) generation
and user editing of an empirical-standards checklist, (4) venue-specific,
formal, and annotated AI reviews, and (5) citation suggestion followed by
reference verification.}
\end{figure*}

\ATIBA{} combines a Django REST Framework backend with a frontend implemented
in Next.js, React, and TypeScript. Checks run asynchronously, and manuscript
results and venue/track data are cached for reuse. The service uses
GPT-5.4\footnote{OpenAI GPT-5.4 model: \url{https://openai.com/index/introducing-gpt-5-4/}}
through Azure OpenAI. Figure~\ref{fig:pipeline} summarizes the five
pipelines.

\subsection{Reference-Integrity Check}
\label{sec:hallucination}

\paragraph{Motivation.} A fabricated or misattributed reference is one of the
most concrete, checkable forms of research misconduct, yet verifying dozens
of references by hand against bibliographic databases is time-consuming
and repetitive.

\paragraph{Design.} \ATIBA{} extracts the manuscript text and hands the PDF
to \texttt{hallucinator-cli}~\cite{hallucinatorcli}, an external tool invoked
as a subprocess, which verifies each reference
against bibliographic sources and returns one of five per-reference statuses:
\texttt{verified} (a matching record was found and the cited authors match),
\texttt{not\_found} (no matching record was found), \texttt{author\_mismatch}
(a matching record was found but the cited authors differ), \texttt{retracted}
(the matching work has been formally retracted), or \texttt{skipped} (the
citation could not be parsed, or the lookup timed out). With URL matching
enabled (\texttt{--url-match}), unresolved references carrying a URL are also
checked for liveness.

Because bibliographic verification establishes only that a reference
\emph{exists}, not that it is used appropriately, \ATIBA{} complements it
with a citation-context check. For each numbered reference, the system
locates bracketed in-text citation markers, including citations appearing in
groups, and extracts up to three sentences in which the reference is cited.
Azure OpenAI then compares these citing contexts with the reference title and
assesses whether the cited work plausibly supports, exemplifies, motivates,
or is otherwise directly related to the claims being made. It returns a
reference-level verdict of \emph{relevant}, \emph{irrelevant}, or
\emph{uncertain}, together with a short rationale. The
\emph{uncertain} verdict is used when the citing context or available
bibliographic metadata is insufficient for a reliable judgment. As shown in Figure~\ref{fig:pipeline}
(pipeline~1), this contextual assessment complements the bibliographic
verdict in the reference-integrity report: a reference may therefore be
verified as a real publication while still being flagged as potentially
misapplied in the manuscript. Results are cached and exportable as a PDF
report, with colour-coded statuses and per-reference details such as
cited-versus-found author names, DOI or arXiv information, retraction notices,
the extracted citation contexts, and the rationale for any contextual
mismatch.

\paragraph{Advantage.} Splitting existence-verification (delegated to a
dedicated, independently maintained tool) from context-relevance judgment
(where an LLM's ability to read intent is actually useful) means \ATIBA{}
never asks the LLM to answer a question a database lookup can answer more
reliably, and reserves the LLM step for the one judgment a lookup cannot
make.

\subsection{Venue/Track Compliance Check}
\label{sec:venue}

\paragraph{Motivation.} Every venue states its own submission rules --
templates, page limits, anonymization policy, reference-desk-reject
conditions -- on its own call-for-papers page, in its own words, and authors
routinely miss a rule stated on a track-specific page rather than the venue's
general homepage.

\paragraph{Design.} Given a track's stored link (or a manually supplied
call-for-papers URL), \ATIBA{} scrapes that specific page using BeautifulSoup~\cite{beautifulsoup} and runs two separate extraction passes over an LLM: one for peer-review-rubric criteria (novelty, rigor, relevance), one for submission/format criteria observable
from the PDF itself (page limits, template, anonymization, reference-integrity
policy, open-science expectations). Several major publishers -- the ACM
Digital Library, Elsevier/ScienceDirect, IEEE Xplore -- sit behind bot
detection and reject automated requests outright, so \ATIBA{} distinguishes
``the page could not be fetched'' from ``the page loaded but stated no
criteria'': in the former case it asks the user to paste the
call-for-papers text in directly and runs the same extraction passes over
that instead. Each extracted item is required to carry
a \emph{verbatim} quote from the scraped page as evidence and a source URL;
items without a locatable quote are not returned. Before the
manuscript-level pass, every extracted criterion first passes through an
evidence gate: the user can edit the extracted requirement's wording, accept
it as-is, or reject a criterion that does not actually apply to their
submission (Figure~\ref{fig:pipeline}, pipeline 2). A third pass then checks
each accepted criterion against the manuscript text and page count,
returning a per-criterion result -- criterion, requirement, observed value,
pass/fail, and the originating evidence quote -- not a single aggregate pass/fail verdict.

\paragraph{Advantage.} Deriving criteria from the venue's own page, not from a fixed internal template, lets the same engine serve venues with
very different rules without per-venue engineering, and the evidence-quote
requirement ensures that only criteria stated in the source are presented.
The evidence gate
additionally keeps a human in the loop before any criterion is enforced, so
an extraction error can be corrected before it silently fails the manuscript.

\subsection{Empirical-Standards Checklist}
\label{sec:empirical}

\paragraph{Motivation.} The ACM SIGSOFT Empirical Standards give the
community a shared definition of what a rigorous study of a given
methodology should report~\cite{ralph2020empirical}, but applying an
eighteen-methodology checklist by hand to every submission does not scale.

\paragraph{Design.} \ATIBA{} ships three checklist catalogues derived from
the standards: a pre-submission checklist for authors, and one-phase and
two-phase review checklists for reviewers at conferences and journals
respectively. Against the selected catalogue, \ATIBA{} first asks an LLM which of its
eighteen methodologies (e.g.\ controlled experiment, case study, repository
mining) the manuscript uses, together with a catch-all entry for studies that
fit none of them and a one-sentence rationale for each selection. For each
selected methodology, the corresponding checklist items are grouped into
essential, desirable, and extraordinary attributes. A second pass then assigns
a \emph{yes}/\emph{partial}/\emph{no} status, together with a justification and
a short evidence quote. Every evidence quote is checked for verbatim presence in the
manuscript; a quote that cannot be located is discarded, while the
verdict and justification are kept, and a \emph{no} verdict with no
justification is automatically softened to \emph{partial}, so it never
stands as an unsupported hard fail. The resulting checklist is then presented to
the user for review: verdicts, justifications, and evidence quotes can be
edited before the checklist is finalized (Figure~\ref{fig:pipeline},
pipeline 3), so the automatically generated checklist remains a draft for the author or reviewer to confirm, and is never treated as a final judgment.

\paragraph{Advantage.} The hallucination defence means an ungrounded
\emph{no} -- the verdict most likely to be disputed by an author -- cannot
survive on invented evidence, which matters most exactly where a wrong
verdict would be most damaging.

\subsection{AI Review}
\label{sec:review}

\paragraph{Motivation.} Different reviewing moments call for different
review shapes: a venue-specific read that adapts to a venue's own rubric, a
fixed, comparably-structured review useful for calibration across
submissions, and a page-anchored review useful for revising a specific
passage.

\paragraph{Design.} \ATIBA{} offers three modes from one venue-aware
prompting layer, all backed by Azure OpenAI~\cite{azureopenai}
(Figure~\ref{fig:pipeline}, pipeline 4). The \emph{venue-specific} mode
produces a decision, contribution summary, critical review, prioritized
improvement list, and scope analysis, substituting the venue's own extracted
review criteria (Section~\ref{sec:venue}) in place of a fixed rubric when
they are available. The \emph{formal review} mode keeps the same structure
across venues -- the venue and track are named in the prompt, but the
sections and the scoring scheme do not change: a summary, strengths,
weaknesses, section-level comments, questions for authors, an overall
evaluation and a
categorical recommendation from strong reject to strong accept, with an
explicit instruction not to invent missing content. The \emph{annotated
review} mode processes the manuscript page by page and returns
quote-anchored annotations -- each tagged strength, weakness, or suggestion,
with a severity and a verbatim quote from that page -- alongside its own
scores (soundness, presentation, contribution, and reviewer confidence) and
recommendation, rendered as highlights directly on the PDF.

\paragraph{Advantage.} Sharing one prompting layer across three modes means
a venue's extracted criteria (Section~\ref{sec:venue}) automatically improve
the venue-specific review without separate integration work, while the formal
and annotated modes give two different, independently useful views onto the
same underlying model call.

\subsection{Citation Recommendation}
\label{sec:citationsuggestion}

\paragraph{Motivation.} Authors often know a claim needs a citation but not
which one, and manually finding a fitting, real, non-fabricated reference is
exactly the kind of task where a plausible-sounding but false suggestion is
most damaging -- a hallucinated \emph{suggested} citation is no safer than a
hallucinated \emph{existing} one.

\paragraph{Design.} Given a passage or claim in the manuscript, \ATIBA{}
asks an LLM to propose candidate citations. Each one is looked up against three bibliographic
sources -- CrossRef, Semantic Scholar, and OpenAlex -- before it is shown
(Figure~\ref{fig:pipeline}, pipeline 5). Only a candidate that a source
affirmatively \emph{verifies} reaches the user; a candidate no source can
confirm -- whether because it does not exist, because the sources disagree
with the cited year, or because a lookup failed -- is discarded; it is never surfaced with a caveat.

\paragraph{Advantage.} Applying the same principle as the
reference-integrity check (Section~\ref{sec:hallucination}) -- a citation is
believed because a bibliographic record says so, never because the model
asserted it -- means citation suggestion cannot become a second, unguarded
path for the same fabrication problem the rest of \ATIBA{} is built to
catch. A proposed citation is held to the same evidentiary bar as one
already in the manuscript.

\section{Evaluation}
\label{sec:evaluation}

Before running the formal survey, we conducted an initial internal deployment by running ATIBA's reference-integrity check on six manuscripts from the authors' research group. Out of the 413 references processed, the check surfaced several citations for manual inspection, ultimately confirming seven distinct publication-year mismatches distributed across three of the manuscripts.

\subsection{User Study}
\label{sec:study}

To complement this initial deployment experience, we conducted a moderated user study to assess the perceived usefulness of the broader system. Participants used the publicly deployed instance of \ATIBA{} to review their manuscripts and subsequently completed a post-task survey.

\paragraph{Participants.} 13 non-author participants were recruited from
the authors' network via direct invitation: five undergraduate students, five
master's students, two PhD students and one research staff member. Four
reported less than one year of research experience, seven reported one to two
years, and two reported six to ten years.

Self-reported effort gives a baseline for the manual work \ATIBA{} targets.
Reviewing a paper took most participants three hours or more (10 of 13,
with three to five hours the most common answer), but checking that paper's
references took markedly less: 11 of 13 spent at most two hours on it, 6 of
them less than one hour, and only 2 participants reported spending longer on
references than on the rest of the review. The reference-integrity check
therefore addresses the part of reviewing to which our participants devote the
least time -- which is also the part where an unverified citation is most
likely to survive.

\begin{figure}[t]
\centering
\includegraphics[width=\columnwidth]{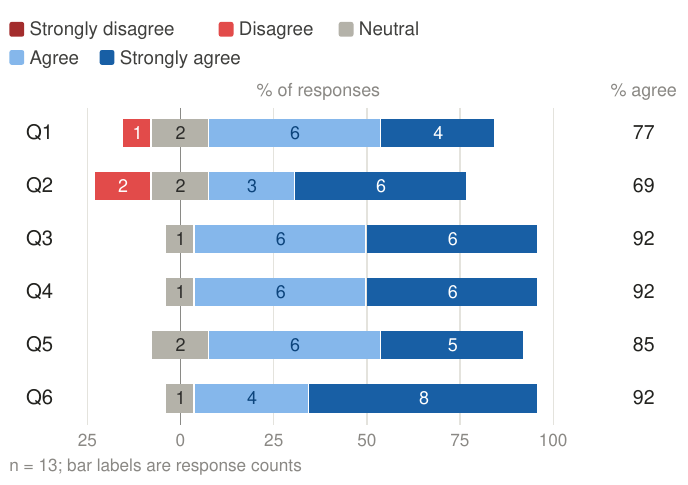}
\caption{Participant agreement with Q1--Q6 ($N=13$), centred on neutral; bar labels show response counts and \% agree combines Agree and Strongly agree.}
\label{fig:likert}
\Description{Diverging stacked bar chart of six survey items. Bars are
centred on the neutral response, with disagreement extending to the left and
agreement to the right. Adoption intent, the venue-criteria check and the
AI-generated review each reach 92 percent agreement, the empirical checklist
85 percent, surfacing overlooked issues 77 percent, and the hallucination
check 69 percent.}
\end{figure}

\subsubsection{Procedure and Instrument}

Each participant used \ATIBA{} on their own manuscript, running the
reference-integrity (hallucination) check, the venue/track compliance check,
the empirical-standards checklist, and the citation-suggestion feature, and
reading the resulting AI review. Participants' target venues
for the venue/track compliance check included top software-engineering
venues such as FSE, ICSE, and ASE.
They then completed a survey. The instrument
first collected the position, experience and effort items reported above, and
then asked the six five-point
agree/disagree items below, quoted verbatim from the instrument and referred
to as Q1--Q6 throughout this section and in Figure~\ref{fig:likert}.

\begin{enumerate}[label=\textbf{Q\arabic*.},leftmargin=2.1em,labelsep=0.4em,
itemsep=1pt,topsep=3pt,parsep=0pt,partopsep=0pt]
\item The system helped me identify potential quality or integrity issues in
the manuscript that I might otherwise have overlooked.
\item The hallucination check helped me identify and verify potentially
incorrect, unverifiable, or unjustified references through clear explanations.
\item The venue criteria check helped identify whether the manuscript complied
with the requirements of the selected venue and track.
\item The AI-generated review accurately identified relevant strengths and
weaknesses of the manuscript.
\item The empirical checklist helped identify missing or insufficiently
reported methodological information in the paper.
\item I would use this system as part of my future paper-writing or internal
peer-review workflow.
\end{enumerate}

The instrument closed with an explicit consent question; responses are
reported anonymously.

\subsubsection{Results}
\label{sec:results}

All 13 retained participants consented to the inclusion of their responses.
Figure~\ref{fig:likert} reports agreement with each item.

Agreement was high across all six items, with means between 4.0 and 4.5, and
no participant selected \emph{strongly disagree} anywhere in the instrument.
The strongest item was adoption intent: 12 of 13 participants (92\%) agreed or
strongly agreed that they would use \ATIBA{} in their future paper-writing or
internal peer-review workflow (Q6, $M=4.5$), eight of them strongly. The
venue-criteria check and AI-generated review each reached the same 92\%
agreement (Q3 and Q4, both $M=4.4$), and the empirical-standards checklist
followed at 85\% (Q5, $M=4.2$).

For Q1, 77\% agreed that \ATIBA{} surfaced quality or integrity issues they
might otherwise overlook ($M=4.0$), with one disagreement and two neutral
responses. The reference-integrity check drew 69\% agreement (Q2, $M=4.0$,
$SD=1.2$), with two disagreements and two neutral responses. Thus, all six
items received majority agreement, with the greatest response variation on
reference integrity.

\section{Threats to Validity}

\textbf{Internal validity.}
The study is a moderated, single-session evaluation and measures participants' perceived usefulness rather than objective accuracy. Moderator presence and the absence of controlled task comparisons may have influenced their responses.

\textbf{External validity.}
The evaluation involved 13 non-author participants from the authors' research network,
most with up to two years of research experience. The findings may therefore
not generalize to more experienced researchers, reviewers, program chairs, or
authors from other disciplines and institutions. Moreover, several \ATIBA{}
features rely on a particular LLM deployment, prompts, and configuration.
Their behavior may vary across model providers, model versions, and future
updates, limiting the generalizability and reproducibility of the observed
results. Broader participant samples, multiple manuscript domains and LLM
configurations, and controlled task-based evaluations are needed to establish
generalizability and accuracy more firmly.

\section{Limitations and Future Work}
\label{sec:limitations}

\textbf{Evaluation coverage.} Our study establishes perceived usefulness in a
moderated setting. Measuring precision and recall on labeled manuscripts with
known reference and compliance defects is the next step toward quantifying
check accuracy. A controlled benchmark can vary the number and type of seeded
reference defects, separating tool accuracy from the particular manuscript
used in the moderated study.

\textbf{Venue-source coverage.} Venue compliance is derived from the textual
content available on the supplied page. Requirements contained only in linked
PDFs, images, or separate documents are outside that input. When a page cannot
be fetched, users can paste its text and run the same evidence-grounded
pipeline. This fallback is useful for publisher pages that reject automated
requests, but it still relies on the supplied text containing the applicable
rule. Extending collection across linked materials would broaden coverage.

\textbf{LLM judgment.} Grounding makes verdicts auditable by tying them to
bibliographic records or source quotations, but context relevance,
rubric-based review, and empirical-standard fulfillment still require model
judgment. The displayed evidence and editable criteria keep these judgments
open to user verification.

\textbf{Reference verification.} Bibliographic results depend on database
coverage and on \texttt{hallucinator-cli}~\cite{hallucinatorcli}'s extraction
of reference titles; we observed occasional title-extraction errors on
IEEE-formatted references and database timeouts, both of which can yield a
\emph{not\_found} result for a legitimate publication. The interface therefore
presents \emph{not\_found} as a prompt for inspection rather than proof of
fabrication; future evaluation will quantify these cases and their effect on
false-positive rates.

\textbf{Citation suggestion.} Verification establishes that a suggested work
exists and matches its bibliographic metadata; it does not by itself establish
that the work is the best support for a particular claim. Evaluating
claim--citation relevance is therefore an additional direction for the
suggestion pipeline.

\textbf{Manuscript confidentiality.} LLM-backed checks process manuscript
text, or page images for annotated review, through Azure OpenAI, whose policy
retains prompts and completions for up to 30 days for abuse monitoring rather
than model training~\cite{azureopenaiprivacy}. A locally hosted processing
option would extend \ATIBA{} to settings with stricter data-residency
requirements.

\section{Conclusion}

Manual manuscript checking remains a repetitive and fragmented part of the
publication process, particularly for reference verification, venue-specific
requirements, and reporting standards. \ATIBA{} brings these tasks together
through five complementary workflows: reference-integrity verification,
venue/track compliance checking, an empirical-standards checklist, multi-mode
AI review, and verified citation suggestion. Across these workflows, the
system is designed to ground model judgments in independently checkable
evidence and to keep criteria, evidence, and outputs visible to the user. Our moderated study provides initial evidence that researchers find this
approach useful, though broader evaluations with labeled defects and more
diverse participants are still needed to establish objective accuracy.
\ATIBA{} does not aim to replace authors or reviewers; it aims to make routine
integrity and quality checks more systematic, transparent, and easier to
verify.

\section*{Data Availability}
\ATIBA{}'s source code, the
anonymized user study responses, and a demonstration video are
available in our replication package\footnote{\url{https://doi.org/10.6084/m9.figshare.33235254.v2}}.

\begin{acks}
We thank Teams 3 and 8 for their contributions to the initial version of this project, developed as part of Bilkent University’s CS319 course in Spring 2026.

The authors used OpenAI's ChatGPT during the preparation of this manuscript to assist with drafting, editing, and improving the clarity and readability of the text. All generated content was carefully reviewed, revised, and validated by the authors. The authors take full responsibility for the accuracy, originality, and integrity of the final manuscript.

\end{acks}

\newpage

\bibliographystyle{ACM-Reference-Format}
\bibliography{references}

\end{document}